# Imaging Strain-Localized Single-Photon Emitters in Layered GaSe Below the Diffraction Limit

*Weijun Luo,[1] Benjamin Lawrie,[2,3]\* Alexander Puretzky,[2]\* Qishuo Tan,[1] Hongze Gao,[1] Gage Eichman,[2] Edward Mcgee,[4] Anna K Swan,[4,5] Liangbo Liang,[2] Xi Ling[1,5,6]\**

[1] Department of Chemistry, Boston University, Boston, MA 02215, United States

[2] Center for Nanophase Materials Sciences, Oak Ridge National Laboratory, Oak Ridge, TN 37831, United States

[3] Materials Science and Technology Division, Oak Ridge National Laboratory, Oak Ridge, TN 37831, United States

[4] Department of Electrical Engineering, Boston University, Boston, MA 02215, United States

[5] The Photonics Center, Boston University, Boston, MA 02215, United States

[6] Division of Materials Science and Engineering, Boston University, Boston, MA 02215, United States

\*Correspondence: xiling@bu.edu; lawriebj@ornl.gov; puretzkya@ornl.gov

This manuscript has been authored in part by UT-Battelle, LLC, under contract DE-AC05-00OR22725 with the US Department of Energy (DOE). The US government retains and the publisher, by accepting the article for publication, acknowledges that the US government retains a nonexclusive, paid-up, irrevocable, worldwide license to publish or reproduce the published form of this manuscript, or allow others to do so, for US government purposes. DOE will provide public access to these results of federally sponsored research in accordance with the DOE Public Access Plan (http://energy.gov/downloads/doe-public-access-plan).

ABSTRACT

Nanoscale strain control of exciton funneling is an increasingly critical tool for the scalable production of single photon emitters (SPEs) in two-dimensional materials. However, conventional far-field optical microscopies remain constrained in spatial resolution by the diffraction limit and thus can only provide a limited description of nanoscale strain localization of SPEs. Here, we quantify the effects of nanoscale heterogeneous strain on the energy and brightness of GaSe SPEs on nanopillars with correlative cathodoluminescence, photoluminescence, and atomic force microscopies supported by density functional theory simulations. We report the strain-localized SPEs have a broad range of emission wavelengths from 620 nm to 900 nm. We reveal substantial strain-controlled SPE wavelength tunability over a ~ 100 nm spectral range and two-orders of magnitude enhancement in the SPE brightness at the pillar center due to Type-I exciton funneling. In addition, we show that radiative biexciton cascade processes contribute to the observed CL photon superbunching. Also, the measured GaSe SPE photophysics after electron beam exposure shows the excellent stability of these SPEs. We anticipate this insight into nanoscale strain control of two-dimensional SPEs will guide the development of truly deterministic quantum photonics.

# INTRODUCTION

Over the past decade, the idea of using nanoscale strain gradients to funnel energy into "artificial atoms" has generated substantial excitement[1]. Strain manipulation of continuously varying electronic band energies and optical bandgaps in flexible two-dimensional (2D) layered materials has enabled wide tunability of exciton binding energies and transport dynamics[2]. For instance, excitons in monolayer transition metal dichalcogenide (TMDC) $MoS_2$ experience a redshift of 0.09 eV under a 2.5% localized uniaxial tensile strain[3]; under biaxial tensile strain, they exhibit a linear redshift rate of 0.11 eV/%[4,5]. In addition, dynamic exciton funneling induced by AFM indentation in monolayer $WS_2$ has resulted in strain-dependent exciton-to-trion conversion; while a similar study on $WSe_2$ demonstrated exciton funneling with a 12 meV redshift at ~ 3% biaxial tensile strain due to reduced n-type doping compared with $WS_2$[6].

Meanwhile, interest in exciton funneling has been stimulated by the discovery of single photon emitters (SPEs) associated with defects and strain-confined excitons in 2D materials such as layered hBN[7–9] and monolayer TMDCs (i.e., $WSe_2$)[10–12]. Specifically, exciton funneling by periodically patterned nano-stressors has emerged as a powerful way to generate deterministic SPEs in monolayer TMDCs[13–16] with controlled position, wavelength, purity, and brightness for widespread applications in quantum information science.

Nonetheless, capturing fine details of exciton funneling with diffraction-limited far-field optical probes remains a critical challenge in studies of SPEs in 2D materials. Exciton diffusion lengths in unstrained monolayers[17–19] have been reported to be less than 1 μm, and they can be substantially smaller (i.e., < 200 nm) when exciton funneling is induced by nanoscale strain[14,20]. Near-field characterization tools with spatial resolution below the diffraction limit are increasingly needed to understand these effects. For example, cathodoluminescence (CL)[21–26] microscopy and spectroscopy have been used to study defect-related SPEs in 2D hBN, and tip-enhanced PL[27–29] and scanning tunneling luminescence[30] microscopies have been used to study localized emitters in quantum dots, carbon nanotubes, and monolayer $WSe_2$. CL microscopy provides a unique approach to perform non-contact measurements at low temperatures with exceptional selectivity in identifying specific areas of interest and compatibility with photon antibunching measurements. However, the electron beam-induced damage can be a concern for some materials. Indeed, monolayer TMDCs must be encapsulated with hBN[31–33] for CL imaging, but imaging of these multilayer structures introduces additional challenges because of the potential for electron beam-induced excitations in each layer of the stack[34].

Compared to extensively studied 2D SPE hosts such as monolayer $WSe_2$ and hBN, strain-localized SPEs in multilayer GaSe remain relatively under-explored. Defect-trapped excitons were first explored as SPEs in multilayer GaSe in 2017[35,36], but GaSe SPEs were only recently deterministically generated through strain localization provided by lithographically patterned nanopillar arrays[16].

Herein, we explore the emission characteristics of locally strained multilayer GaSe using CL microscopy and spectroscopy supported by atomic force microscopy (AFM), photoluminescence (PL), and photon antibunching measurements. We show a broad range of emission wavelengths from 620 nm to 900 nm of the strain-localized SPEs. With correlative CL and AFM measurements, we reveal that the strain plays a vital role in controlling the energy and brightness of localized GaSe emitters by observing a continuous redshift in emission wavelengths from 680 nm to 780 nm and enhancement of emitter brightness by two orders of magnitude with increasing strain. The spectral tuning is explained by a strain-induced redshift of the bandgap modeled with density functional theory (DFT) simulations. Moreover, the enhanced emission intensity from the substrate to the pillar apex can be attributed to exciton funneling resulting from the strain-induced decrease in conduction band minimum (CBM) and increase in valence band maximum (VBM). Photon bunching and antibunching of emitters on pillars are observed in CL and PL measurements, respectively, and no beam-induced degradation of the emitter photophysical properties are observed. By combining these correlative characterization techniques, we have gained insights into how nanoscale strain affects the formation, wavelength, and intensity of GaSe SPEs. This research provides a comprehensive understanding of the photophysics of SPEs in multilayer GaSe, sheds light on the potential for engineering SPEs in 2D materials using strain and lays a crucial foundation for developing 2D quantum photonic devices.

## RESULTS AND DISCUSSION

**Hyperspectral CL mapping of multilayer GaSe on SiO$_2$ nanopillars**

To introduce biaxial tensile strain to GaSe, we stacked ε-type GaSe flakes mechanically exfoliated from high-quality bulk crystals[37] onto a Si substrate with lithographically patterned SiO$_2$ nanopillar arrays using the dry-transfer method as described in our previous work[16]. As illustrated in **Figure 1a**, the flake pins to the upper edges of each nanopillar forming a circular biaxially tented region. This circular tent forms a strain field where the biaxial tensile strain starts at zero outside the circle and then increases continuously until maximizing at the tent apex. **Figure 1b** shows an SEM image of a thick GaSe flake (thickness ~ 150 nm) tented on SiO$_2$ nanopillar arrays. Hyperspectral CL images are acquired for a variety of nanopillars and demonstrate a wide emission wavelength distribution spanning ~590 nm (~2.10 eV) to ~900 nm (~1.38 eV). Note GaSe is an indirect semiconductor with broadband direct and indirect exciton luminescence between ~590 nm (~2.10 eV) and ~620 nm (~2.0 eV) as a result of the small energy difference between its direct gap (2.07 eV) and indirect gap (2.03 eV)[38]. **Figures 1c and 1d** highlight the large-area CL response across the inset optical image in **Figure 1b**. **Figure 1c** illustrates the integrated CL intensity for wavelengths between 590 nm (~ 2.10 eV) and ~ 620 nm (~ 2.0 eV), corresponding to the band edge emission of GaSe; in this band, most of the CL is observed around the circumference of the tents, with minimal CL observed at the center of the pillars or in the unstrained areas between pillars. **Figure 1d** illustrates the integrated CL intensities between 650 and 900 nm below the indirect bandgap of unstrained GaSe. These sub-bandgap emissions localize near the centers of the tents. The full hyperspectral CL response across these pillars is presented in **Supplementary Video I**. A statistical description of the emission wavelengths of the strongest emission band in each CL spectrum is summarized and shown in **Supplementary Figure S1a.**

While the spectrum image shown in **Figure 1** provides a global picture of the effect of strain on excitons in GaSe, it is limited to 180 nm spatial resolution because of the 100 ms dwell time required to acquire a spectrum at each pixel and the size of the composite image. In the following sections, we investigate the strain-controlled emission properties with higher spatial resolution and reduced field of view.

**Hyperspectral CL line scan and AFM measurements of multilayer GaSe on $SiO_2$ nanopillars**

In order to better understand the effects of nanoscale strain gradients on excitons in GaSe, a high-resolution CL line scan across two pillars with a spatial resolution of 20 nm was performed on the 70-nm-thick flake shown in **Figure 2a**. **Figure 2b** shows an AFM height profile across pillars A and B. The fabricated pillars have conical frustum geometries as a result of the reactive ion etching process, and the full width at half maximum (FWHM) of pillars A and B are 450 nm and 410 nm, respectively. When GaSe is transferred on the pillars, the average height of both tents is ~ 100 nm while the average FWHM is ~ 1.1 µm. The average distance between both ends (diameter) of a tent is ~ 2.8 µm.

The dashed white arrow in the SEM image in the left panel of **Figure 2c** illustrates the CL line-scan path from the region slightly above tent A (position 1), across the apex of tent A (position 2), the flat region between tent A and B (position 3), and the apex of tent B (position 4), to the flat region below tent B (position 5). The right panel of **Figure 2c** shows the position-dependent CL spectra acquired by the CL line scan. As the electron beam approaches the apex of tents A and B, the broad band-edge emission between 590 nm (~ 2.10 eV) and 620 nm (~2.0 eV) experiences a slight redshift (up to ~ 20 nm), and its intensity decreases gradually. This observation of decreasing indirect-exciton emission intensity with increasing strain is common in 2D layered materials[39,40]. At the tent apex, the band edge emission almost vanishes. Meanwhile, sub-bandgap CL bands near 700 nm (~1.77 eV) and 750 nm (~1.65 eV) are strongly enhanced and redshifted with increasing strain. We attribute these bands to the localized excitons that are associated with trap states in GaSe[35,38,41–43]. In addition, as shown in **Figure S2a**, the sub-bandgap CL enhancement was quantified through the position-dependent integrated CL intensity ratio $\frac{I_{X_S}}{I_{E_g}}$, which is the ratio of the integrated sub-bandgap CL intensity ($I_{X_S}$, between 650 nm and 800 nm) over the band-edge CL intensity ($I_{E_g}$, between 590 nm and 650 nm). The sub-bandgap CL is two orders ($\frac{I_{X_S}}{I_{E_g}} = 184$) and one order ($\frac{I_{X_S}}{I_{E_g}} = 16$) of magnitude stronger than the band-edge CL on tents A and B. The CL results for tents A and B are almost the same except for the sub-bandgap emission intensity, which implies that the sub-bandgap CL bands on each pillar originate from the same trap states with different carrier concentrations.

Compared to the band-edge emission centered at 610 nm in the unstrained area, the sub-bandgap CL bands on the pillar apexes are significantly redshifted (up to ~ 100 nm). Similar strain-induced redshifts and enhancement of emission intensities have been reported for SPEs in monolayer WSe$_2$ that are strained by nanopillars[13,14]. The increasing strain from the pillar edge to pillar apex results in a continuous decrease in the bandgap and tuning of the bandgap across randomly distributed trap states, resulting in significantly improved radiative efficiency[14,15]. Moreover, a previous study on PL emission of unstrained GaSe flakes[44] suggests that the prevalence of non-radiative decay pathways due to surface traps causes the low quantum efficiency of the sub-bandgap luminescence.

Five CL spectra collected at positions 1 – 5 are presented in **Figure 2d**: spectra 1 and 3 (collected at the edges of tents A and B) show weak band edge CL centered at 610 nm (~ 2.03 eV) and even weaker sub-bandgap CL bands near 700 nm (1.77 eV); spectra 2 and 4 are collected at the apexes of tents A and B, where the sub-bandgap emission at ~ 780 nm dominates the CL spectrum; spectrum 5 is collected away from tents A and B and only shows weak band-edge CL centered at 605 nm (2.05 eV) and 610 nm (~ 2.03 eV) without any sub-bandgap CL observed.

**Correlative microscopy and strain simulations of a single tent**

To understand the spatial correlation between nanoscale biaxial strain and excitonic luminescence in a quantitative fashion, we perform hyperspectral CL mapping of a single tent with ~40 nm spatial resolution and reveal its correlation with the biaxial strain within that tent based on Landau continuum theory[20,45]. **Figure 3a** shows an SEM image of tent A (highlighted in the optical image of **Figure 2a)**, where the dark square indicates the CL mapping area. No beam-induced damage was observed within this area, but a higher pixel density was used within the CL mapping area, causing the change in contrast. **Figure 3b** shows an AFM image of the same area. The area in the dashed circle is the tent area, which is ~ 1.3 µm in diameter. **Figure 3c** illustrates the simulated strain divided into three regions by the two dashed circles: 1. high-strain area (1.0% - 2.6%); 2. low-strain area (0% - 1.0%); 3. strain-free area (0%). The strain-free region is more than ~ 1.4 µm away from the center of tent A; the high-strain area (1.0% - 2.6%) is less than ~ 0.6 µm from the center of tent A; the low-strain area lies in the annulus between the strain-free and high-strain areas.

This delineation of distinct strain regions is enabled by the analysis of the principal components of the spectrum image via non-negative matrix factorization (NMF). NMF is a powerful linear algebra algorithm for multivariate analysis that has been widely used for modeling spatial and temporal variation and evolution of principal spectral components in spectroscopic studies, including fluorescence images[46,47] and CL microscopy[24,48]. NMF treats a spectrum image as a product matrix M of two matrices W (weight) and P (principal component): M = W×P. Therefore, a CL hyperspectral map can be interpreted as the spatially weighted sum of its spectral components (emission wavelengths). For the CL spectrum image of tent A, ten principal emission lines are identified (see more details in Table S1 in Supplementary information). Here, because initial analysis indicated no strong correlations between distinct spectral lines, the ten spectral components from the CL mapping of pillar A were fit to Voigt functions and used as priors for NMF analysis (see more details in Section 2 of Supplementary Information). The spatial distribution of the three most prominent components centered at 610 nm, 703 nm, and 760 nm are illustrated in **Figure 3d-f**. Other spectral lines with lower intensity are described in supplementary **Figure S3**. **Figure 3d** shows that the band-edge emission centered at 610 nm emerges in the low-strain area between the outer and inner dashed circles. In contrast, sub-bandgap CL at 703 nm and 760 nm is observed primarily in the high-strain area inside the inner circle (**Figures 3e and 3f**). The intensity of the sub-bandgap CL bands in the high-strain area are one order of magnitude stronger than the band edge CL observed in the low-strain area, which is supported by the mapping position-dependent integrated CL intensity ratio $\frac{I_{X_S}}{I_{E_g}}$ shown in **Figure S2b**.

Similar analyses for all ten CL spectral components are shown in supplementary **Figure S4**, clearly illustrating the CL redshift with increasing strain. We also performed PL mapping and NMF analysis over pillar A. The result is summarized in **Figures S5, S6, and S7** and is roughly consistent with the CL results despite the relatively coarse spatial resolution of PL microscopy. In short, the CL mapping results clearly show a strong correlation between the strain and the emission in GaSe, where the band edge emission and sub-bandgap emission wavelengths can be tuned in a large range from ~590 nm (~2.10 eV) – ~ 630 nm (~1.97 eV) and ~ 680 nm (~ 1.82 eV) – ~ 780 nm (~ 1.59 eV), respectively. Moreover, the tensile strain-controlled redshift rate of sub-bandgap emission energy is calculated as ~100 meV/%, which is much larger than that of 51 meV/% for strain-localized WSe$_2$ SPEs[14].

**DFT calculations of band structures of biaxially tensile-strained GaSe**

To better understand the origins of the strain-induced redshifts and changes in the intensities of both band-edge and sub-bandgap CL, we simulate the band structure of bulk GaSe under the biaxial tensile strain of 0% to 6% with DFT using the PBE functional. As shown in **Figure 4a**, under increasing biaxial tensile strain, the conduction band energies continuously decrease while the valence band energies increase (Supplementary Figure **S8a**), resulting in a bandgap redshift rate of -0.16 eV/% (**Figure 4b**). While DFT calculations with PBE functionals usually underestimate bandgaps, our focus here is on the change in bandgap with increasing strain rather than on quantitative models of the actual bandgap. Thus, the observed redshift of band-edge CL is consistent with the strain-induced bandgap redshift from the simulation results.

GaSe retains its indirect gap across the full range of modeled strain; no strain-induced indirect-to-direct transitions are observed from the simulation results. However, the DFT model does show that the momentum difference between the CBM and VBM for the strained GaSe is larger than that of unstrained GaSe (Supplementary **Figure S8b**), which might explain the strain-induced decrease in band edge CL intensity. In addition, the decreasing emission intensities of indirect excitons with increasing strain are common in 2D layered materials[39,40].

Moreover, the strain-induced redshift of bandgaps also contributes to the observed redshift in the sub-bandgap CL. Based on the modeled redshift rate of -0.16 eV/%, the strain map in **Figure 3d** can be replotted as the strain-redshift map shown in **Figure 4c**. The increasing strain from the edge to the apex of the tent results in decreasing bandgaps, and the maximum amount of biaxial tensile strain of ~ 2.6% at the apex of the tent corresponds to a CL redshift to ~770 nm, consistent with the maximum observed CL emission wavelength of 780 nm. **Figure 4d** further illustrates the strain-dependent energies of the CBM and VBM; the position-dependent energy states of the CBM and VBM exhibit funnel shapes consistent with "Type-I" funneling[1], where the CBM decreases and the VBM increases with increasing strain.

"Type-I" funneling yields bandgap gradients that result in the aggregation of excitons approaching the funnel center[49] and the formation of a "2D exciton reservoir"[50]. The exciton drift length is a critical metric of the effective formation of a 2D exciton reservoir. The effect of the exciton drift length on the number of excitons collected at the funnel center has been extensively studied in 2D materials[1,4,50–52]. In order to realize efficient exciton funneling to the center, the exciton drift length $l_{drift}$ should be set at least to be the funnel radius R (defined as the distance from the unstrained area to the center), $l_{drift}=R$. The exciton drift lengths of GaSe excitons confined by nanoscale strain can be described as[1]:

$$l_{drift} = \langle v \rangle_{drift} \tau_{1/2} = \frac{\nabla E_{exc}}{m_{exc}} \tau_{dephase} \tau_{1/2}$$

Where $\tau_{1/2}$ is the exciton lifetime, $\tau_{dephase}$ is the phase relaxation time, and $\langle v \rangle_{drift}$ is the average drift velocity. $\nabla E_{exc}$ is the spatial exciton energy gradient and $m_{exc}$ is the exciton mass. Therefore, $\nabla E_{exc}$ can be approximated by $\frac{\Delta E_{exc}}{R}$, leading to the radius of the funnel R:

$$R = \sqrt{m_{exc}^{-1} \Delta E_{exc} \tau_{dephase} \tau_{1/2}}$$

In the unstrained region, the energy variation $\Delta E_{exc}$ of GaSe excitons is ~50 meV[38]. Since the exciton mass of GaSe is not well described in the literature, we use the electron rest mass $m_e$ ($0.511 \times 10^6$ eV) to represent the exciton mass $m_{exc}$, and the dephasing time $\tau_{dephase}$ of GaSe exciton is ~11 ps[53]. The lifetime $\tau_{1/2}$ of GaSe direct excitons has been reported as ~40-200 ps at 4.2 K[44,54]. Then, the exciton drift length in the unstrained region is calculated as ~1.9–4.4 µ$m$ and larger than the tent radius of ~1.4µ$m$. Thus, excitons can be efficiently funneled to the center of the pillar tent.

Besides, the defect and impurity[55,56] existing in GaSe introduce trap states[44] lying below the band gap. Bound excitons that associated with these trapped states overlap in the emission spectrum and exhibit as broad peaks[35,43]. Previous works have identified that pressure[57] and strain gradients[14,15] can segregate these trap states and introduce discrete energy states. Intriguingly, the discrete energy states residing in the strain-induced energy funnel show discrete lines strong high intensity counts on the emission spectrum similar to those of quantum wells and quantum dots[58]. Thus, the probability of exciton formation is increased while the non-radiative recombination pathways[59] are decreased, resulting in a higher probability of emitting photons upon excitation. To summarize, the increasing strain not only creates a "Type-I" funnel by red-shifting the bandgaps but also segregates trap states. The combined effect leads to the observation of remarkably enhanced and well-isolated sub-bandgap emission peaks.

**Photon statistics of SPEs under electron-beam and optical excitation**

Finally, we performed Hanbury Brown–Twiss (HBT) measurements on the CL and PL from the same pillar region in order to understand the photon correlations associated with the strain localized exciton manifold. **Figure 5a** illustrates the CL spectrum of an emitter on pillar A, and **Figure 5b** shows the associated photon correlation function with $g^{(2)}(0) = 2.92\pm0.03$ with a time constant of 4.2 ns (fitting details are discussed in Supplementary Section **8**). The observed CL bunching is consistent with past reports of the impulsive high-energy electron excitation of ensembles of emitters[21,25,26]. **Figure 5c** shows a PL spectrum acquired on pillar A after the CL measurements with an excitation power density of 1 nW/µ$m^2$, and this 627 nm emitter was also observed. However, the PL emission shows weak antibunching with $g^{(2)}(0) = 0.59\pm0.06$, and a time constant of 0.48 ns (**Figure 5d)** in contrast to the bunching and slower dynamics observed in CL. The 627 nm emission from the PL measurement can be deconvolved into an exciton peak at 1.980 eV (626.2 nm) and two biexciton peaks at 1.979 eV (626.6 nm) and 1.981 eV (625.9 nm) according to the high-resolution PL spectrum shown in **Figure S9a**. The spectral overlap between the exciton and biexciton features is clearly detrimental to the GaSe exciton SPE purity.[16] Indeed, as shown in **Figure S9c**, given a larger excitation power density of 10 nW/µ$m^2$, the purity of this SPE continues to degrade to $g^{(2)}(0) = 0.72\pm0.07$, and a shelving state[60,61] related to the interaction between exciton and biexciton states emerges with $g^{(2)}(\tau)>1$ for small $|\tau| > 0$. The appearance of the shelving state is a result of the increasing impact of radiative biexcitons on the photon statistics with growing incident power density. With this framework in mind, the CL bunching observed here can be explained as a result of a radiative biexciton cascade[62–65].

Moreover, we characterized another SPE on pillar A at 692 nm (as shown in **Figures 5e and 5f**) by PL excitation and observed $g^{(2)}(0)=0.43\pm0.04$ and a decay time-constant of ~ 0.2±0.02 ns (using a 128 ps time bin). This emitter can also be deconvolved into an exciton peak at 1.792 eV (692 nm) and a biexciton peak at 1.787 eV (694 nm) as described and identified by power-dependent measurements shown in supplementary **Figure S10**. Also, the power-dependent PL antibunching of this SPE is summarized in **Figures S11a** and **S11b.** Notably neither emitter exhibited any changes in photophysics induced by the electron-beam.

CONCLUSION

In conclusion, correlative cathodoluminescence, photoluminescence, and atomic force microscopies complemented by DFT modeling provide a true nanoscale picture of strain-localized SPEs in GaSe. The strain-localized SPEs are with a broad range of emission wavelengths from 620 nm to 900 nm. We show substantial strain-controlled tuning of localized excitons across a ~ 100 nm bandwidth. A "type-I" energy funneling effect leads to luminescence enhancement and redshift with increasing strain. With higher exciton densities at the funnel center, radiative recombination from exciton complexes (excitons and biexcitons) that are associated with trap states are significantly enhanced compared to those in the unstrained area. CL photon bunching with a 4.2 ns time constant was observed and attributed to radiative biexciton cascade processes. Despite the challenge of probing CL antibunching in excitonic complexes, our PL measurements on the same pillar confirm the presence of isolated SPEs. The observation of PL antibunching after CL measurements also highlights the stability of these GaSe SPEs. The combined nanoscale understanding of strain-induced SPE dynamics and energetics provided by these correlative microscopies provides a new framework for the deterministic generation of bright and robust GaSe SPEs with highly controllable and tunable wavelengths. More broadly, a nanoscale understanding of strain and exciton funneling should translate to many other classes of SPEs in 2D materials and provide critical insights for other photonic applications such as lasing, photovoltaic conversion, and optical detection.

**Methods.** *Synthesis of bulk GaSe crystal.* Bulk GaSe was synthesized using the chemical vapor transport (CVT) method[9] as described in our previous work.[16]

*Photoluminescence spectroscopy.* The cryo-PL and associated photon statistics measurements were performed in a home-built confocal PL microscope in a backscattering configuration. A Princeton Instruments Isoplane SCT-320 spectrograph with a PIXIS 400BR Excelon camera and a grating turret with 150 g/mm, 600 g/mm, and 2400 g/mm gratings were used to measure PL spectra with spectral resolutions of ~ 3.25 meV, ~ 300 μeV, and ~ 30 μeV, respectively. A 532 nm diode laser (Cobolt) was used for excitation. A 100x in-vacuum objective (Zeiss, NA = 0.85) was integrated with the Montana S100 closed-cycle cryostat. The PL mapping was controlled by 2-axis galvo scanning. The photon-antibunching measurements utilized a pair of large-area superconducting nanowire single-photon detectors (SNSPDs, Quantum Opus) and a Swabian Time Tagger 20 time-correlated single photon counting (TCSPC) system. A 90:10 non-polarizing beam splitter was used to allow for PL (10% coupling efficiency) and photon correlation functions (90% coupling efficiency) to be acquired in parallel.

*Cathodoluminescence microscopy.* The cathodoluminescence microscopies were performed in an FEI Quattro SEM with a Gatan dual fuel cryostage and a Delmic Sparc cathodoluminescence module. The sample was cooled to 9 K with liquid helium, and all spectra were acquired on an Andor Kymera spectrograph with an Andor Newton CCD camera. All CL photon statistics measurements were performed by fiber coupling the CL into the same SNSPDs used for the photoluminescence measurements.

*DFT simulations.* Plane-wave DFT calculations were carried out using the Vienna Ab initio Simulation Package[28,29] (VASP) with projector augmented wave (PAW) pseudopotentials[28,30,31] for electron-ion interactions, and the generalized gradient approximation (GGA) functional of Perdew, Burke and Ernzerhof[66] (PBE) for exchange-correlation interactions. The details of geometry optimization of bulk GaSe was shown in our previous work.[16] The electronic band structure post-analysis was carried out using the VASPKIT package.[67]

## AUTHOR INFORMATION

**Corresponding Authors**

*Email: xiling@bu.edu; puretzkya@ornl.gov; lawriebj@ornl.gov

The authors declare no competing financial interest.

## ACKNOWLEDGEMENTS

This material is based upon work supported by the National Science Foundation (NSF) under Grant No. (1945364). Work by X.L. was supported by the U.S. Department of Energy (DOE), Office of Science, Basic Energy Sciences (BES) under Award DE-SC0021064. The CL microscopy was supported by the U.S. Department of Energy, Office of Science, National Quantum Information Science Research Centers, Quantum Science Center. The PL and CL spectroscopies were supported by the Center for Nanophase Materials Sciences (CNMS), which is a US Department of Energy Office of Science User Facility. X.L. and A.K.S. acknowledge the membership of the Photonics Center at Boston University. The computational work is performed using Shared Computing Cluster (BUSCC) at Boston University.

## AUTHOR CONTRIBUTIONS

W.L., B.L., A.P., and X.L. conceived the experiment. Q.T. synthesized and characterized the bulk GaSe crystals. W.L. prepared samples with assistance from Q.T., and H.G. B.L. performed the CL microscopy and CL photon statistics measurements. W.L. conducted the PL / photon-statistics measurements with assistance from A.P. and B.L. W.L performed theoretical calculations with assistance from L.L. W.L performed the analysis and interpretation of the data with assistance from A.P., B.L., G.E., E.M., L.L., A.K.S., and X.L. All authors contributed to the writing of the manuscript.

**ASSOCIATED CONTENTS**

**SUPPORTING INFORMATION**

The Supporting Information is available free of charge at XXX. Description of statistics of maximum emission wavelengths, NMF decomposition of CL mapping results, Power-dependent PL spectra, and Power-dependent photon-antibunching results (PDF).

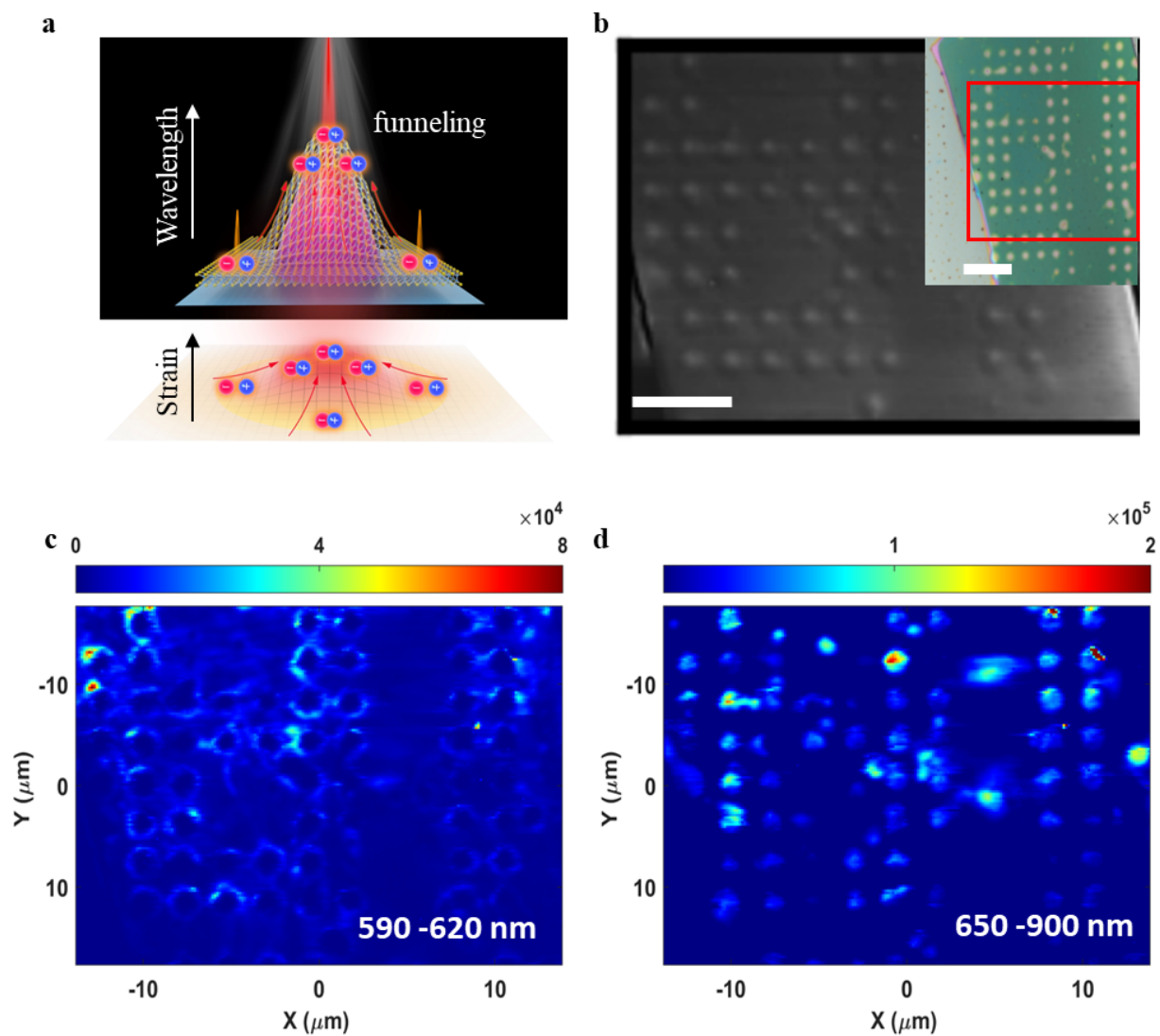

**Figure 1. CL imaging of GaSe flakes on SiO$_2$ nanopillar arrays. (a)** illustration of GaSe on the nanopillar and exciton funneling effect. **(b)** SEM and inset optical images of a 150-nm-thick GaSe flake tented on a SiO$_2$ nanopillar array (scale bar, 10 μm). **(c)** CL intensity map integrated across wavelengths of 590-620 nm for a representative array of pillars (highlighted area in **(b)**) measured at T = 9 K, with an incident electron beam energy of 5 keV, a beam current of 28 pA, and a spatial resolution of 180 nm. **(d)** CL intensity map integrated across wavelengths of 650-900 nm for the same representative array of pillars measured at T = 9 K.

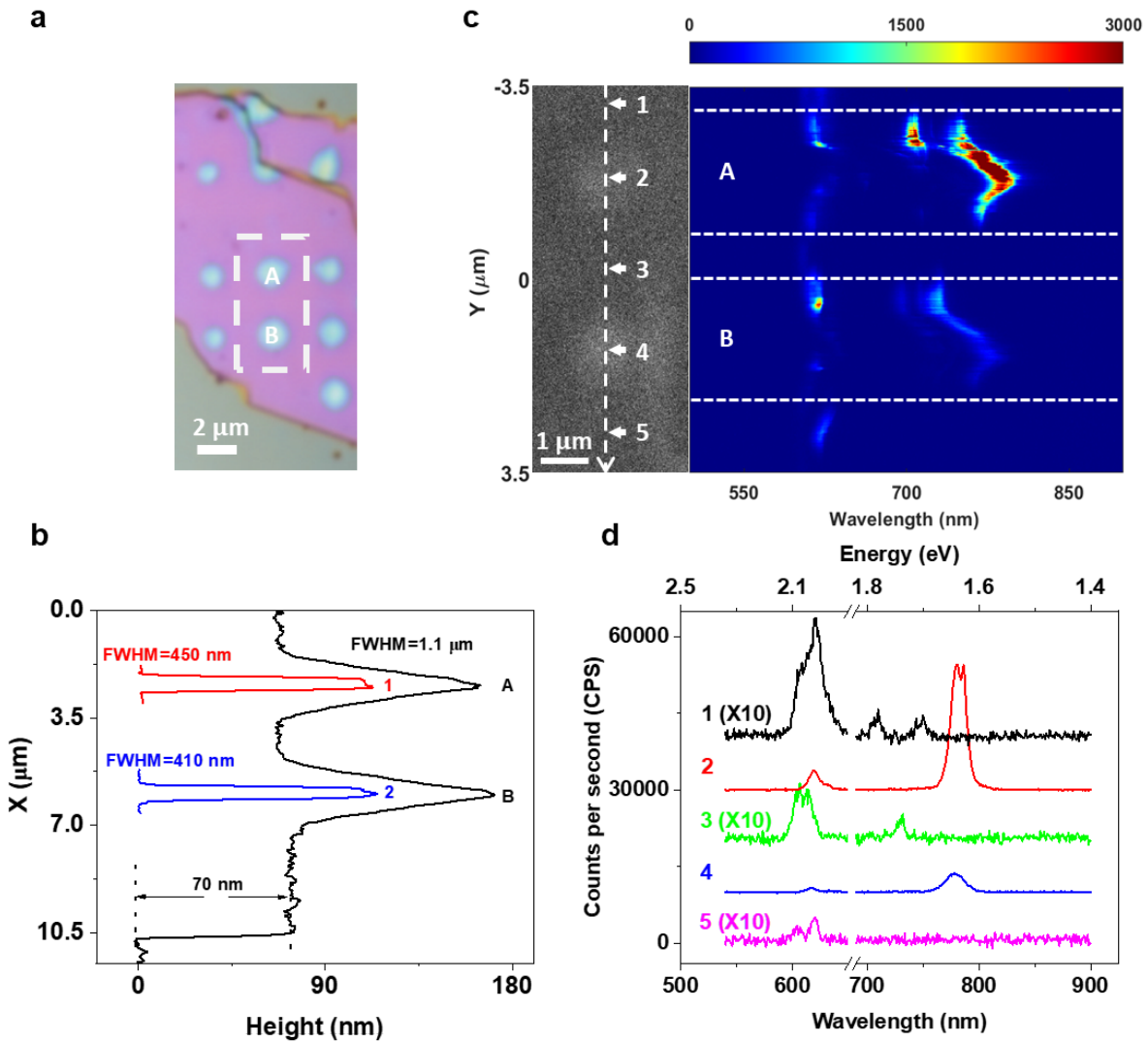

**Figure 2. CL line scan of GaSe flakes on SiO$_2$ nanopillar arrays.** (a) Optical image of a 70-nm-thick GaSe flake tented on a SiO$_2$ nanopillar array (scale bar, 2 μm). CL line scan measurements are performed along Pillars A and B. (b) AFM height profiles of tents A and B (black), and the associated SiO$_2$ pillars A (red) and B (blue). (c) left panel: SEM image of the highlighted area in (a), (scale bar, 1 μm). The dashed white arrow shows the CL line scan path. The right panel shows the CL line scan acquired with 20 nm spatial resolution. Dashed lines are guides to the eye. (d) Five representative CL spectra from positions 1 − 5, as labeled in the left panel of (c). The intensities of spectra 1, 3, and 5 are scaled up 10-fold for comparison with the other spectra.

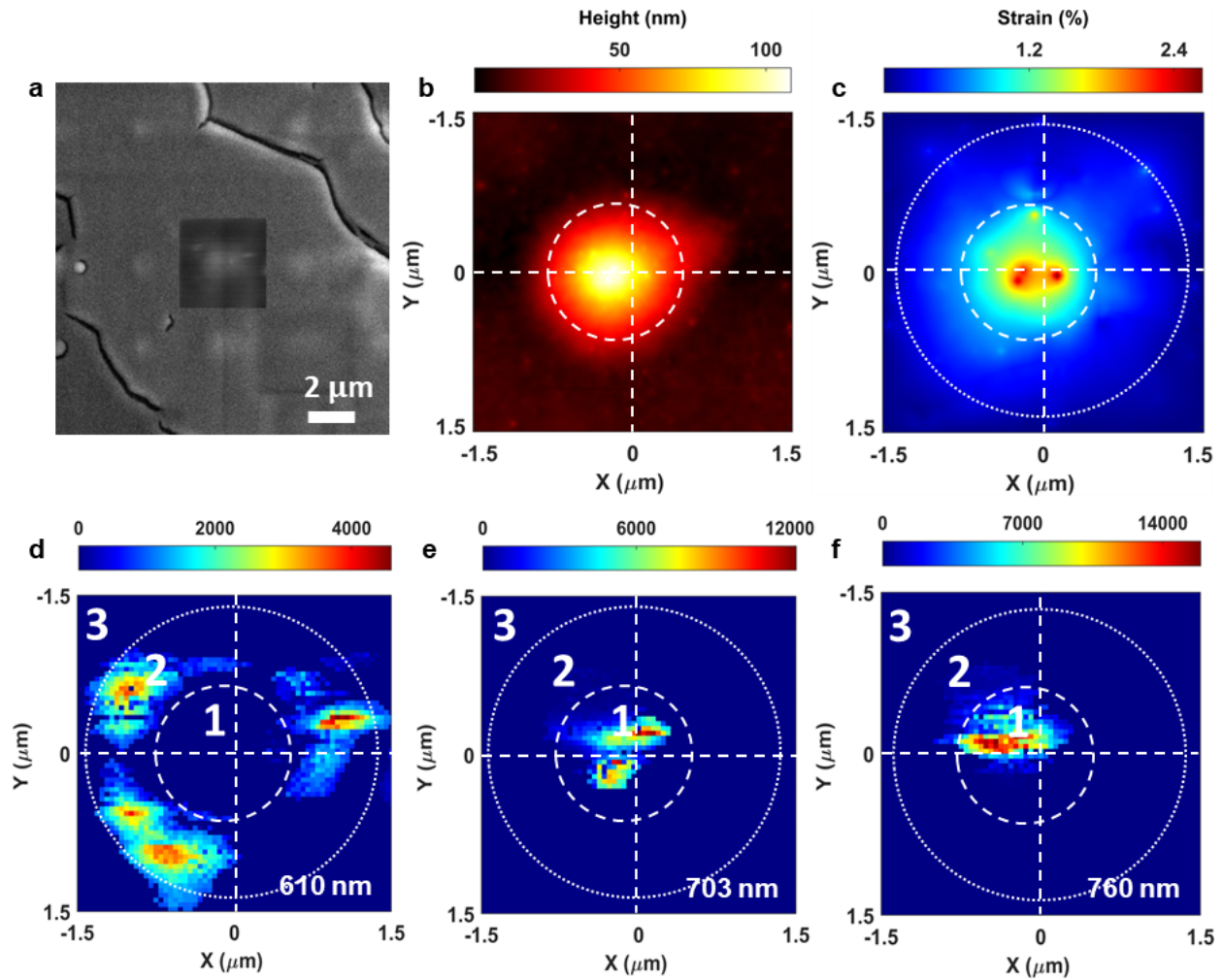

**Figure 3. NMF decomposition of CL spectrum image measured at T=9 K. (a)** SEM image of pillar A (labeled in Figure **2a**) for CL mapping (scale bar, 2 μm). **(b)** AFM image of pillar A**,** dashed lines are guides to the eye. **(c)** Simulated strain distribution over the pillar according to AFM height profile in **(b)**. **(d) − (f)** Intensity maps of three emissions at 610 nm, 703 nm, and 760 nm, respectively.

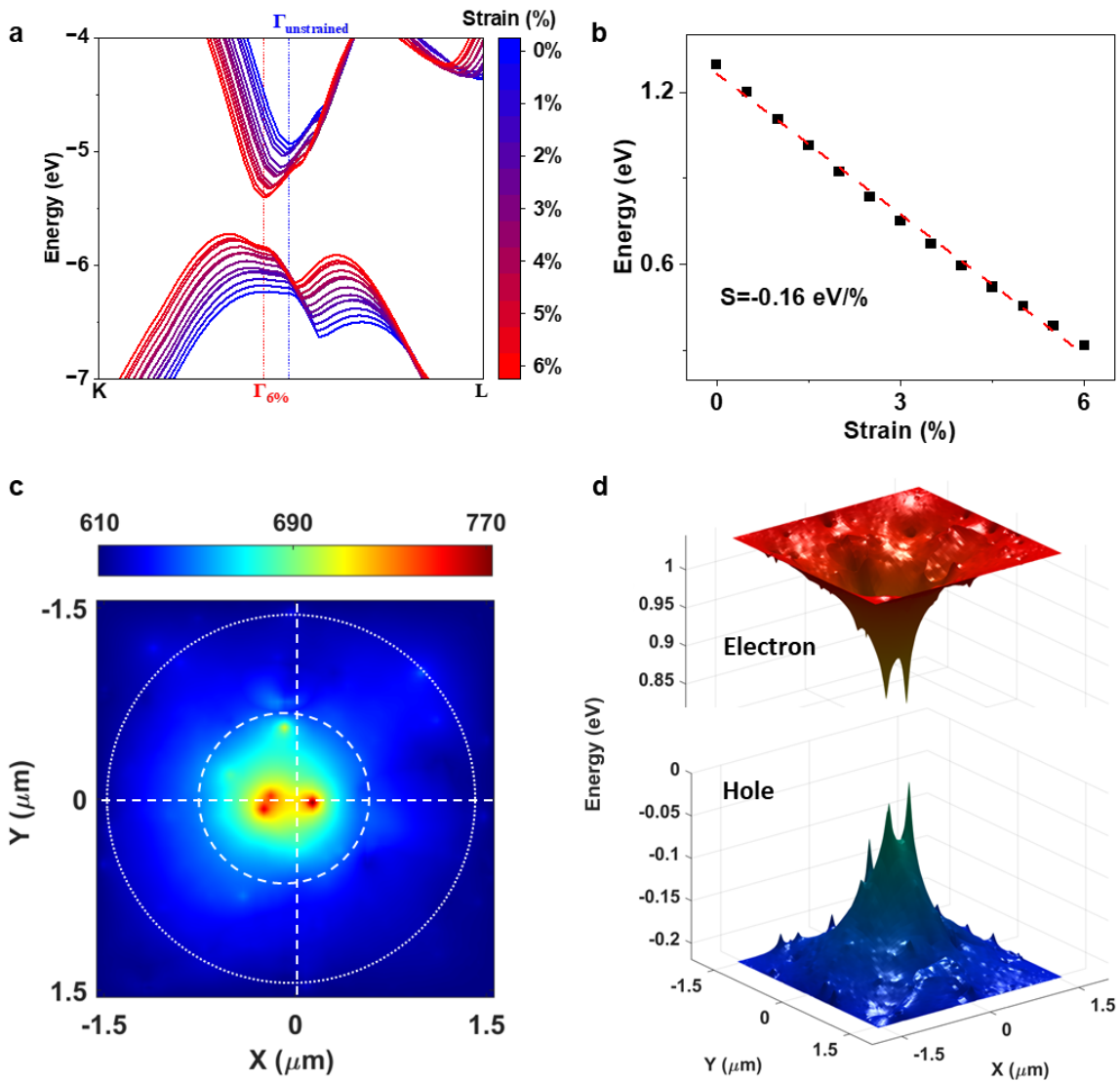

**Figure 4. Biaxial tensile strain-modulated redshift of bandgaps. (a)** Simulated band structure affected by biaxial tensile strain from 0% to 6%. **(b)** Simulated redshift of bandgap (eV) as a function of biaxial tensile strain (%). **(c)** Correlation of the simulated bandgap (eV) with the calculated strain distribution map shown in Figure 2(d), dashed lines are guides to the eye. **(d)** Correlation of the simulated electron (CBM) and hole (VBM) energy profiles with the calculated strain distribution map shown in Figure 2(d).

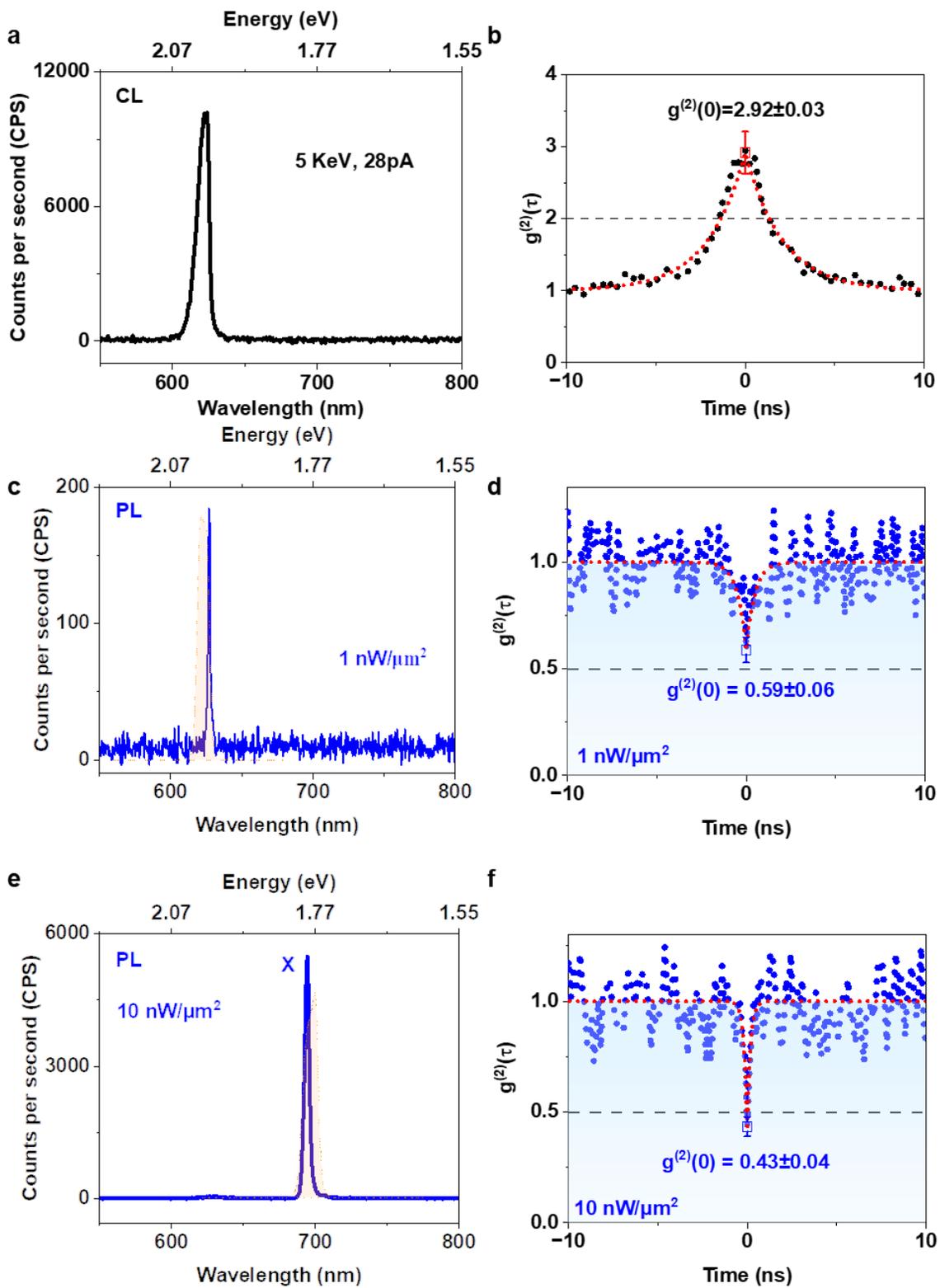

**Figure 5. Comparison of the CL (at 9K) and PL (at 3.5K) photon statistics of emitters at pillar A. (a)** A representative CL spectrum measured on tent A at T = 9 K, with an incident electron beam energy of 5 keV and a beam current of 28 pA. **(b)** Photon bunching measured for this 627 nm emitter acquired without a bandpass filter; the fitted second-order correlation function shows $g^2(0)= 2.92\pm0.03$. **(c)** A PL spectrum collected on tent A at T = 3.5 K and with an incident laser power density of 1 nW/$\mu m^2$, the orange shaded area indicates the spectral window of the narrow bandpass filter for photon antibunching measurements. **(d)** Photon-antibunching measured for this 627 nm emitter acquired using a narrow bandpass filter (625 nm, FWHM) = 10 nm; the fitted second-order correlation function shows $g^2(0)=0.59\pm0.06$. **(e)** Another PL spectrum measured on tent A at T = 3.5 K and with an incident laser power density of 10 nW/$\mu m^2$, the orange shaded area indicates the spectral window of the narrow bandpass filter for photon antibunching measurements. **(f)** Photon-antibunching measured for this 692 nm SPE acquired using a narrow bandpass filter (697 nm, FWHM = 10 nm); the fitted second-order correlation function shows $g^2(0)=0.43\pm0.04$.

Supplementary Materials for

# Imaging Strain-Localized Single-Photon Emitters in Layered GaSe below the Diffraction Limit

*Weijun Luo,[1] Benjamin Lawrie,[2,3]\* Alexander Puretzky,[2]\* Qishuo Tan,[1] Gage Eichman,[2] Edward Mcgee,[4] Anna Swan,[4,5] Liangbo Liang,[2] Xi Ling[1,5,6]\**

[1] Department of Chemistry, Boston University, Boston, MA 02215, United States

[2] Center for Nanophase Materials Sciences, Oak Ridge National Laboratory, Oak Ridge, TN 37831, United States

[3] Materials Science and Technology Division, Oak Ridge National Laboratory, Oak Ridge, TN 37831, United States

[4] Department of Electrical Engineering, Boston University, Boston, MA 02215, United States

[5] The Photonics Center, Boston University, Boston, MA 02215, United States

[6] Division of Materials Science and Engineering, Boston University, Boston, MA 02215, United States

\*Correspondence: xiling@bu.edu; lawriebj@ornl.gov; puretzkya@ornl.gov

**This PDF file includes:**

**Figure S1.** Statistics of maximum emission wavelengths of hyperspectral mapping results

**Figure S2.** Position-dependent CL peak intensity ratio $\frac{I_{X_S}}{I_{E_g}}$

**Figure S3.** NMF decomposition of CL mapping results of tent A in Figure 3 measured at T=9 K: peak areas (integrated intensities).

**Figure S4.** NMF decomposition of CL mapping results of tent A in Figure 3 measured at T=9 K: peak wavelengths.

**Figure S5.** NMF decomposition of PL mapping results of tent A in Figure 3 measured at T=3.5 K: integrated peak intensities.

**Figure S6.** NMF decomposition of PL map of tent A in Figure 5 measured at T=3.5 K: peak wavelengths.

**Figure S7.** Comparison of PL spectra (measured at T=3.5 K) collected on tent A excited with a low and high incident power densities of 10 nW/$\mu m^2$ and 1.19 $\mu W/\mu m^2$, respectively.

**Figure S8.** Supplementary DFT simulation results of Figure 4 in the main text.

**Figure S9.** PL spectra and corresponding antibunching measurements of the 627 nm emitter (measured at T = 3.5 K, incident power density = 1 nW/$\mu m^2$ and 10 nW/$\mu m^2$) shown in the CL spectra in Figure 5a of the main text.

**Figure S10.** Power-dependent PL spectra of the 694 nm (1.787 eV) SPE shown in Figures 5e of main text.

**Figure S11.** Power-dependent photon-antibunching results of the 694 nm (1.787 eV) SPE shown in Figures 5e of main text

Section 1. Statistics of maximum emission wavelengths of hyperspectral mapping results.

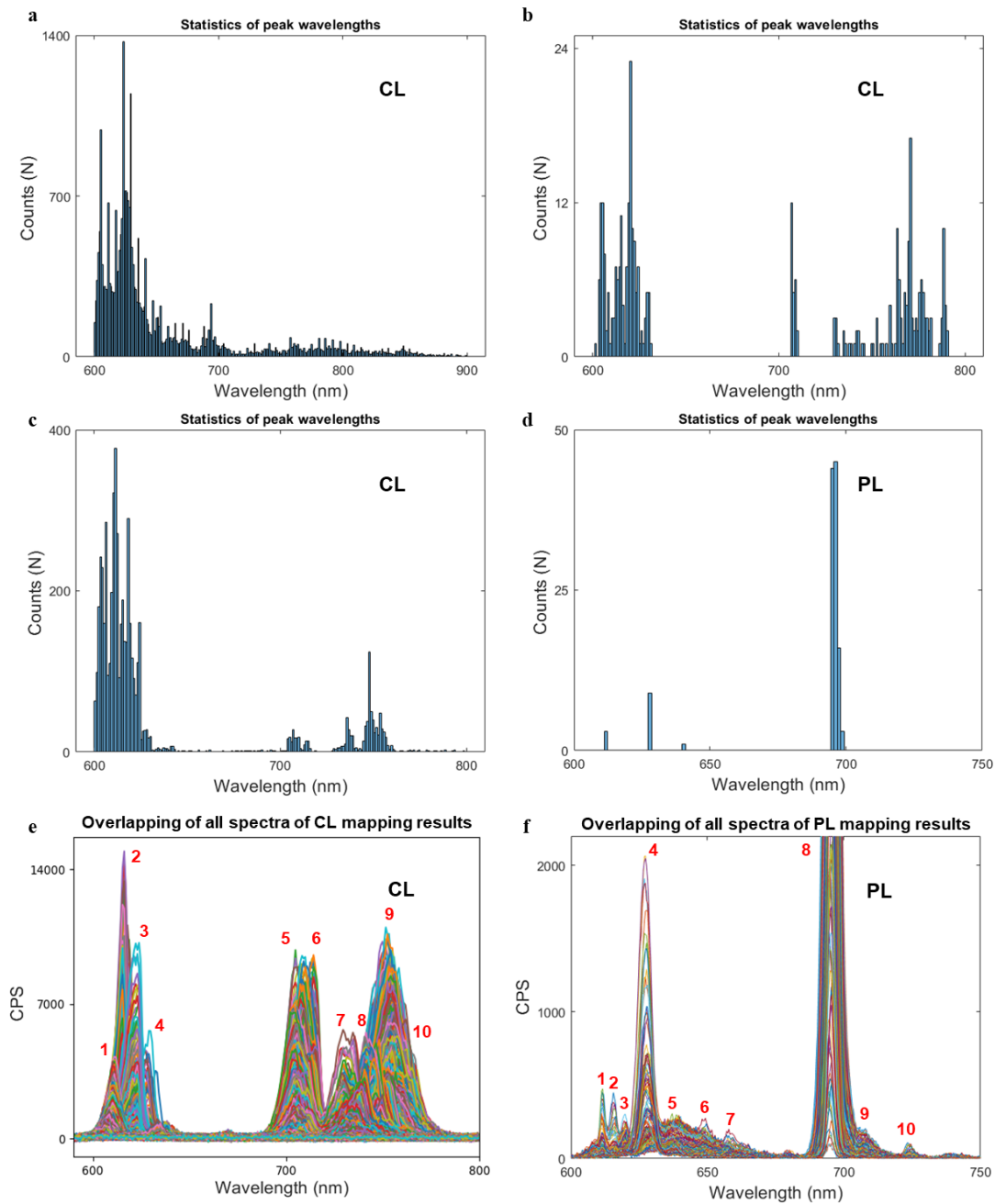

**Supplementary Figure S1. Statistics of** maximum emission wavelengths **of hyperspectral mapping results. (a)** Statistics of maximum emission wavelengths of CL mapping results extracted from Figures **1a** and **b** in the main text. **(b)** Statistics of maximum emission wavelengths of CL line scanning results from Figure **2b** of the main text. **(c)** Statistics of maximum emission wavelengths of CL mapping results of tent A from Figure **3** of the main text. **(d)** Statistics of maximum emission wavelengths of PL mapping results of tent A. **(e)** Overlap all spectra of CL mapping result and label the principal emission wavelengths of tent A. (f) Overlap all spectra of PL mapping result and label the principal emission wavelengths of tent A.

Figures **S1a – c** show the statistics of maximum emission wavelengths of hyperspectral CL mapping results: due to the high spatial resolution (~180 nm, ~20 nm and ~40 nm, respectively), the variation of emission wavelengths affecting by strain at nanoscale can be distinguished easily. However, the maximum emission wavelengths of the PL mapping results of tent A (shown in Figure **S1d**) have less variation due to its relatively lower spatial resolution. The overlapping of all spectra of PL mapping result indicate that there are ten major emission wavelengths that listed in Table S1.

**Section 2. Position-dependent CL peak intensity ratio.**

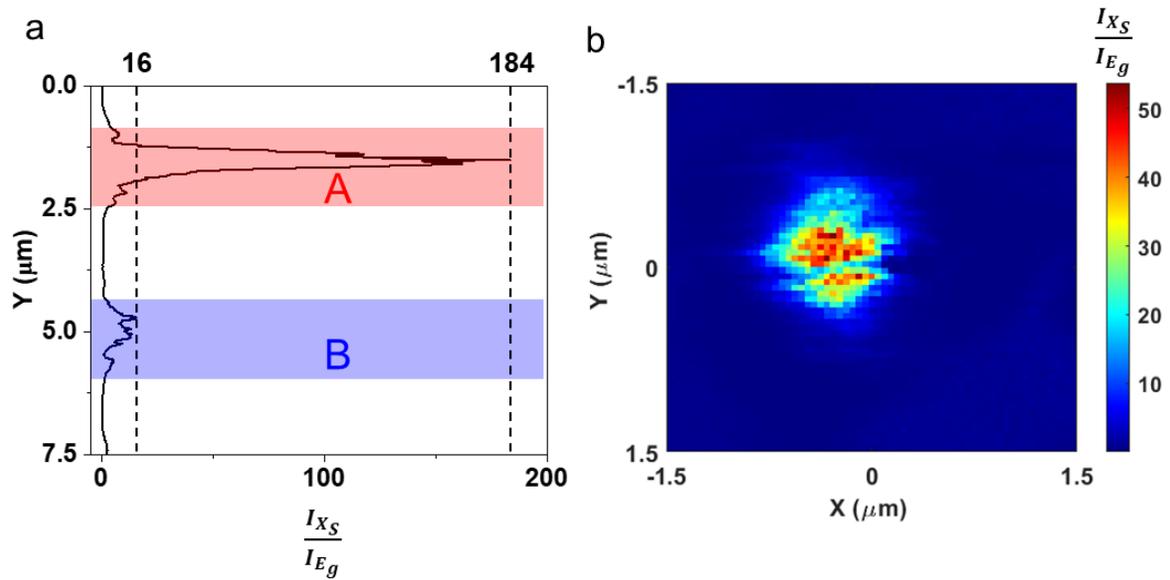

**Supplementary Figure S2.** Position-dependent integrated CL intensity ratio $\frac{I_{X_S}}{I_{E_g}}$ (integrated CL intensity of sub-bandgap emissions ($I_{X_S}$, between 650 nm and 800 nm) over that from the bandedge ($I_{E_g}$, between 600 nm and 650 nm). **(a)** CL scan results of Figure 2c in the main text. **(b)** CL mapping results of Figure 3 in the main text.

Section 3. Non-negative matrix factorization (NMF) of principal spectral components (emission wavelengths): method, implementation and evaluation.

a. Method:

Non-negative matrix factorization (NMF) treats a spectrum image as a product matrix M of two matrices W (weight) and P (principal component): M = W×P. In CL and PL spectra, the principal components are typically the spectral line shapes of the major emission peaks, which contain important information such as wavelengths, intensities, and linewidths. The spectral line shapes of emission peaks are described by Lorentzian and Gaussian functions, respectively due to the homogeneous and inhomogeneous broadening. The Voigt function, which is a convolution of Lorentzian and Gaussian functions, has been extensively adopted for spectral line fitting.[1,2] Hence, the NMF process of a spectrum image can be listed as:

1. Identification of principal spectral components: The first step is to identify the principal spectral components, which are the major contributors to the spectral variation in the data.
2. Fitting of principal spectral components: Once the principal spectral components are identified, they can be fitted to the spectra using Voigt functions. This high-throughput fitting approach enables the simultaneous analysis of all spectra.
3. Extraction and visualization of spectral data: After the fitting is complete, the wavelengths, intensities, and linewidths of the spectral components can be extracted and visualized with position information in the heatmap.

b. Implementation

First, the principal spectral components (emission wavelengths) are screened roughly by the clustering results from the statistics of maximum emission wavelengths obtained from CL and PL mapping results of tent A shown in Figure **S1c** and **S1d.** Then, the principal spectral components (emission wavelengths) are compared with the overlapping of all CL and PL spectra of tent A shown in Figure **S1e** and **S1f.** Ten principal spectral components are identified and listed in Table **S1**.

Here we processed NMF analysis by automated high-throughput spectral fitting via our SmartspecDecov package[3] derived from the "Peak fitting to either Voigt or LogNormal line shapes" package released on Mathworks.[4]

c. Evaluation of fitting

The iterative Voigt fitting of multiple principal components based on least-squares regression was performed for all spectra in both CL and PL mapping results. The least-squares regression can be expressed as:[5]

$$\frac{min\|f(x)\|_2^2}{x} = \frac{min\sum_i (f_i(x)^2)}{x}$$

Here, $f_i(x)$ represents the Voigt function of the i$^{th}$ peak at wavelength x. The coefficient of determination $R^2$ is used for evaluation of how well the regression model (the overall fitted spectrum that consists of the linear combination of multiple principal components) fits the observed data, where $R^2$ can be written as:[6]

$$R^2 = 1 - \frac{RSS}{TSS} = 1 - \frac{\sum_i^n (y_i - \hat{y}_i)^2}{\sum_{i=1}^n (y_i - \bar{y}_i)^2}$$

Where y is the actual data, $\hat{y}_i$ is the fitted data, and $\bar{y}_i$ is the averaged actual data; RSS is the sum of squares of residuals, and TSS is the total sum of squares. The coefficient of determination $R^2$ ranges from 0 to 1, with 0 indicating that the model does not fit the data at all, and 1 indicating a perfect fit. We observe an average $R^2$ value of 0.8 for all 5700 CL spectra.

**Table S1** Summary of ten principal spectral components (emission wavelengths)

|    | PL wavelength (nm) | CL wavelength (nm) |
|----|--------------------|--------------------|
| 1  | 606                | 610                |
| 2  | 614                | 614                |
| 3  | 616                | 621                |
| 4  | 622                | 624                |
| 5  | 631                | 703                |
| 6  | 638                | 707                |
| 7  | 651                | 725                |
| 8  | 694                | 740                |
| 9  | 704                | 750                |
| 10 | 724                | 760                |

Section 3. NMF decomposition of CL mapping results of tent A in Figure 3 of main text

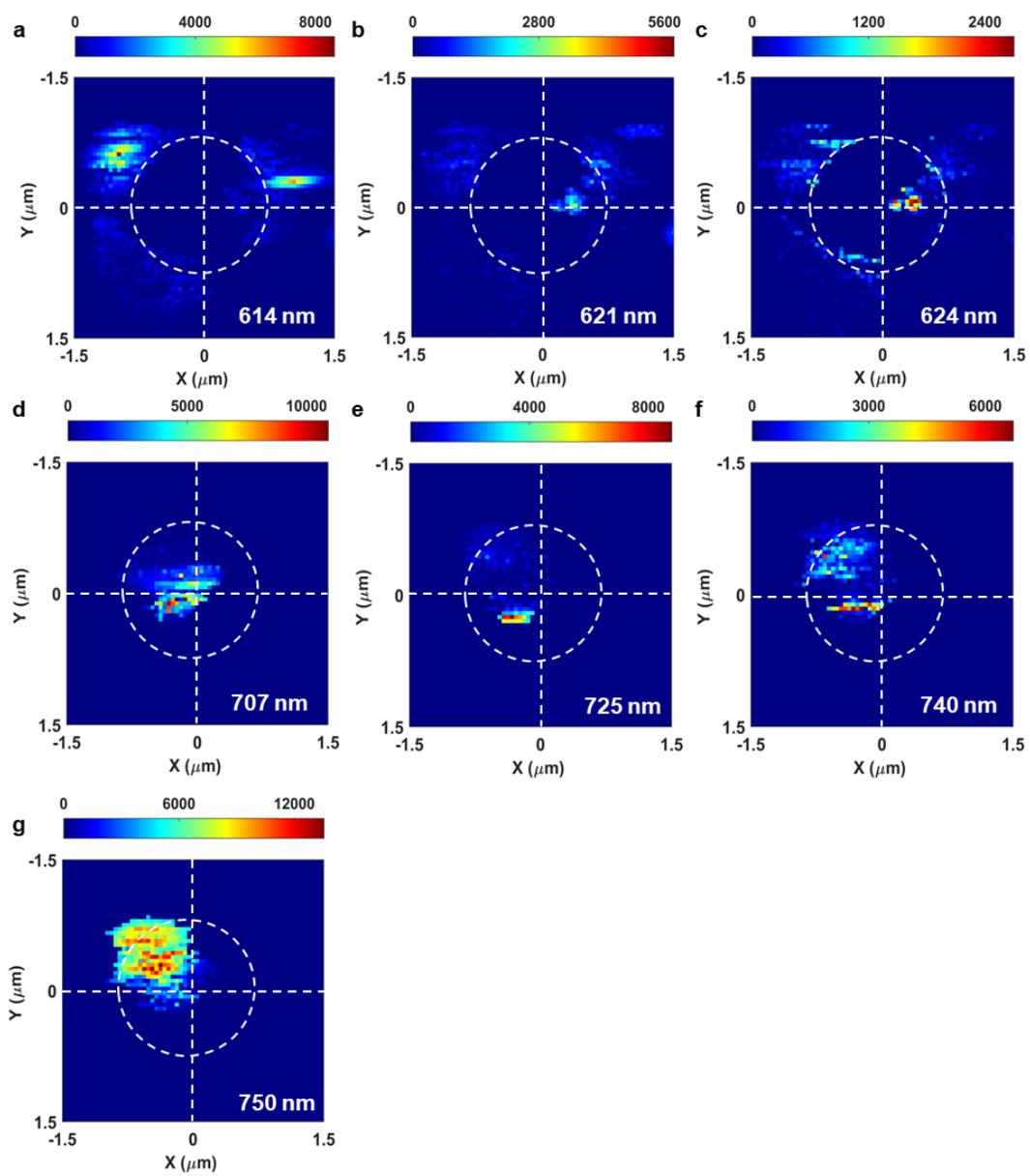

**Supplementary Figure S3.** NMF decomposition of CL mapping results of tent A in Figure 3 measured at T=9 K: peak areas (integrated intensities). **(a) – (c)** Peak areas of band edge CL at 614 nm, 621 nm and 624 nm, respectively. **(d) – (g)** Peak areas of sub-bandgap CL at 707 nm, 725 nm, 740 nm, and 750 nm, respectively.

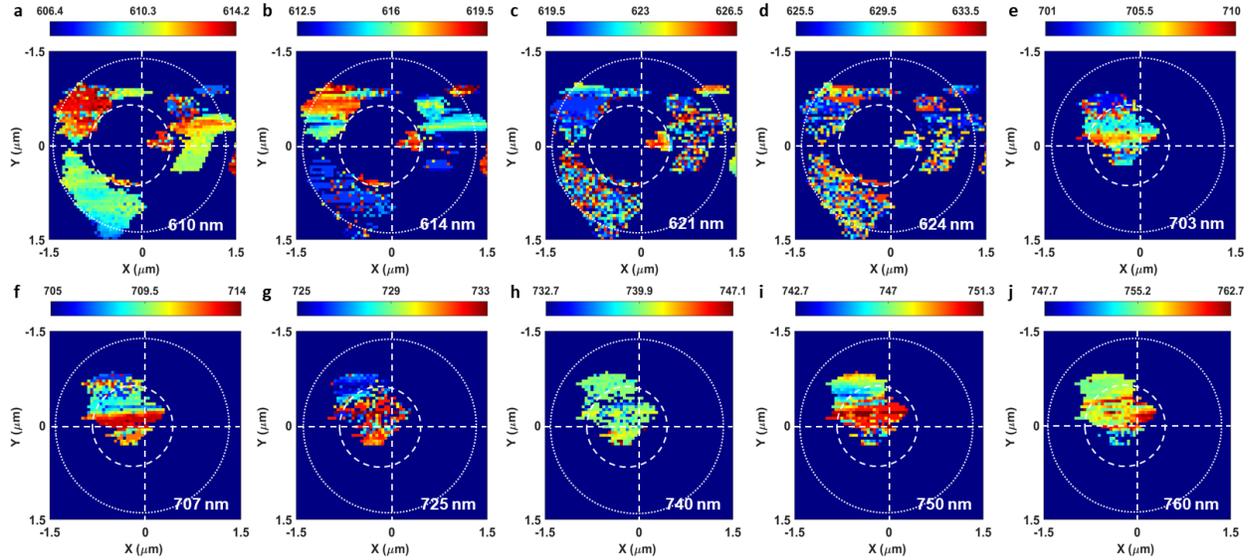

**Supplementary Figure S4.** NMF decomposition of CL mapping results of tent A in Figure 3 measured at T=9 K. **(a) – (d)** Peak wavelengths of band edge emissions at 610 nm, 614 nm, 621 nm and 624 nm, respectively. **(e) – (j)** Peak wavelengths of sub-bandgap emissions at 703 nm, 707 nm, 725 nm, 740 nm, 750 nm, and 760 nm, respectively.

Section 4. Non-negative matrix factorization (NMF) of PL mapping results of tent A.

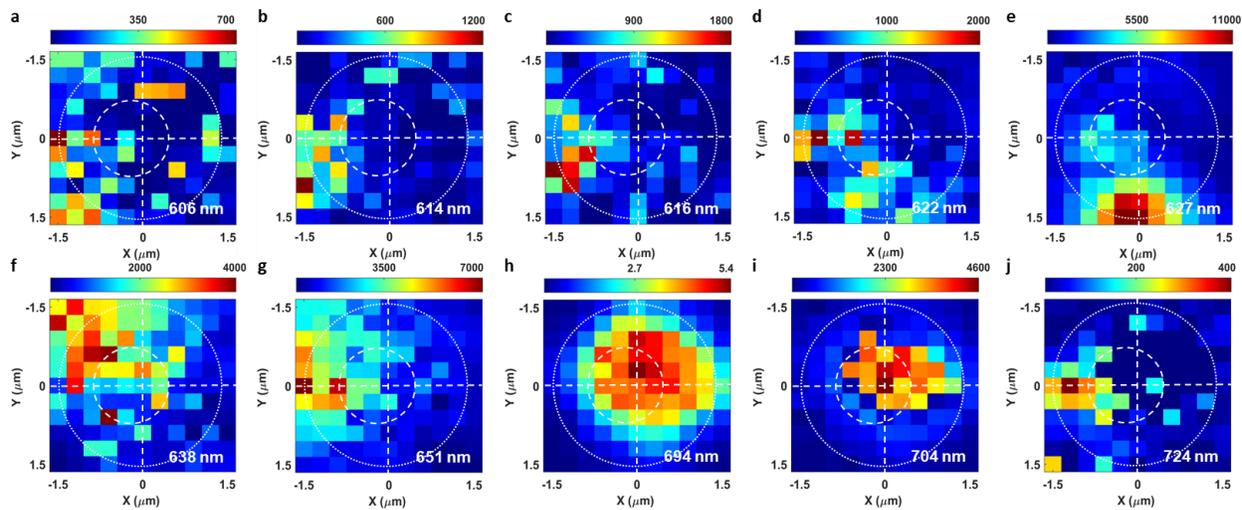

**Supplementary Figure S5 NMF decomposition of PL mapping results of tent A in Figure 3 measured at T=3.5 K. (a) – (j)** Maps of the emission intensities of ten principal components at 606 nm, 614 nm, 616 nm, 622 nm, 627 nm, 638 nm, 651 nm, 694 nm, 704 nm, and 724 nm, respectively.

In order to better understand the GaSe photophysics, a hyperspectral PL map of tent A was acquired at 3.5 K with a low incident power density of 13 nW/μ$m^2$. The NMF decomposition of the hyperspectral PL map (shown in Supplementary **Figure S5**) illustrates ten emission lines from 605 nm to 725 nm; the lower energy 780 nm bands observed with CL are noticeably absent. While the spectral content of the CL maps is independent of bias current and beam energy (changing beam conditions only changed the CL intensity), the spectral content of GaSe PL is strongly dependent on laser power. Additional emission lines with wavelengths longer than 700 nm emerge under higher incident power densities: as shown in **Figure S7**, emission lines at 738 nm and 780 nm emerge under a larger incident power density of 1.19 μW/μ$m^2$, consistent with the observed emission wavelengths from CL results.

Moreover, the band edge PL at 627 nm (as shown in **Figure S5e**) is maximized at the pillar edge and gradually decreases while moving towards the tent apex; in contrast, the sub-bandgap PL at 694 nm (as shown in **Figure S5h**) is maximized at the pillar apex and continuously decreases while moving to the tent edge. Furthermore, the spatial distributions of wavelengths of these two emission lines match well with the emission intensities, which are shown in Supplementary **Figure S6**. Note the laser spot size is about 600 nm, and the step size is ~ 270 nm for this hyperspectral PL map (11 pixels × 11 pixels of 3 μm × 3 μm mapping area), resulting in substantially reduced spatial resolution compared with CL mapping.

Generally, all the luminescent bands exhibit robust spectral stability with no spectral bleaching, blinking and wandering observed in the CL or PL for the beam conditions described here. At higher electron beam currents and dwell times, measurable beam induced damage was observed.

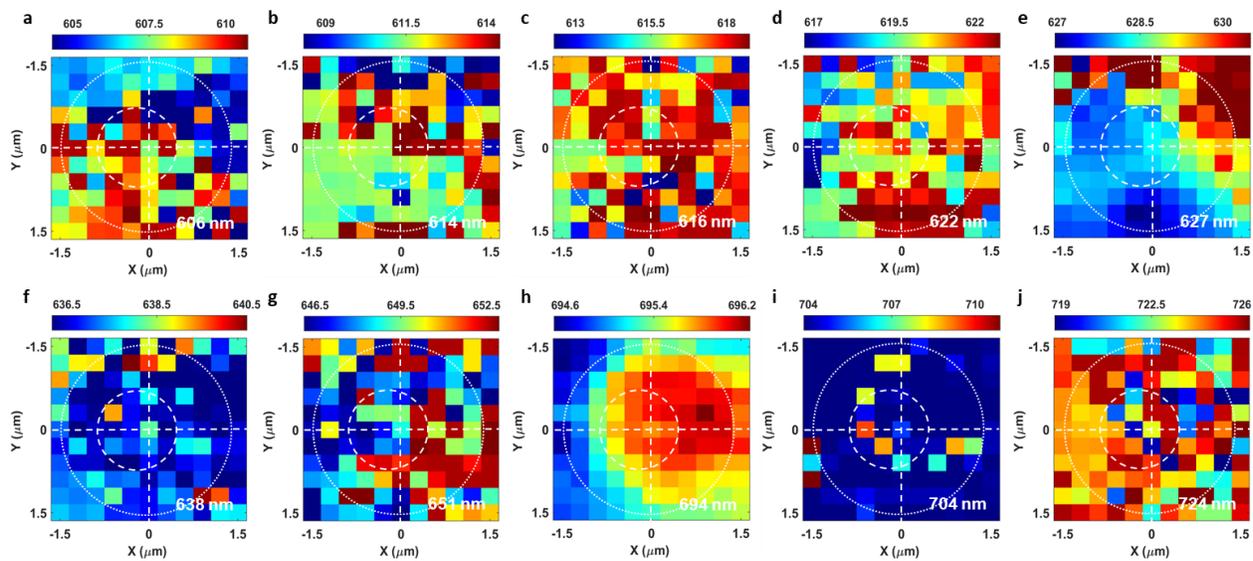

**Supplementary Figure S6 NMF decomposition of PL map of tent A in Figure 5 measured at T=3.5 K: peak wavelengths. (a) – (j)** Peak wavelengths of the PL at 606 nm, 614 nm, 616 nm, 622 nm, 631 nm, 638 nm, 651 nm, 694 nm, 704 nm, and 724 nm, respectively.

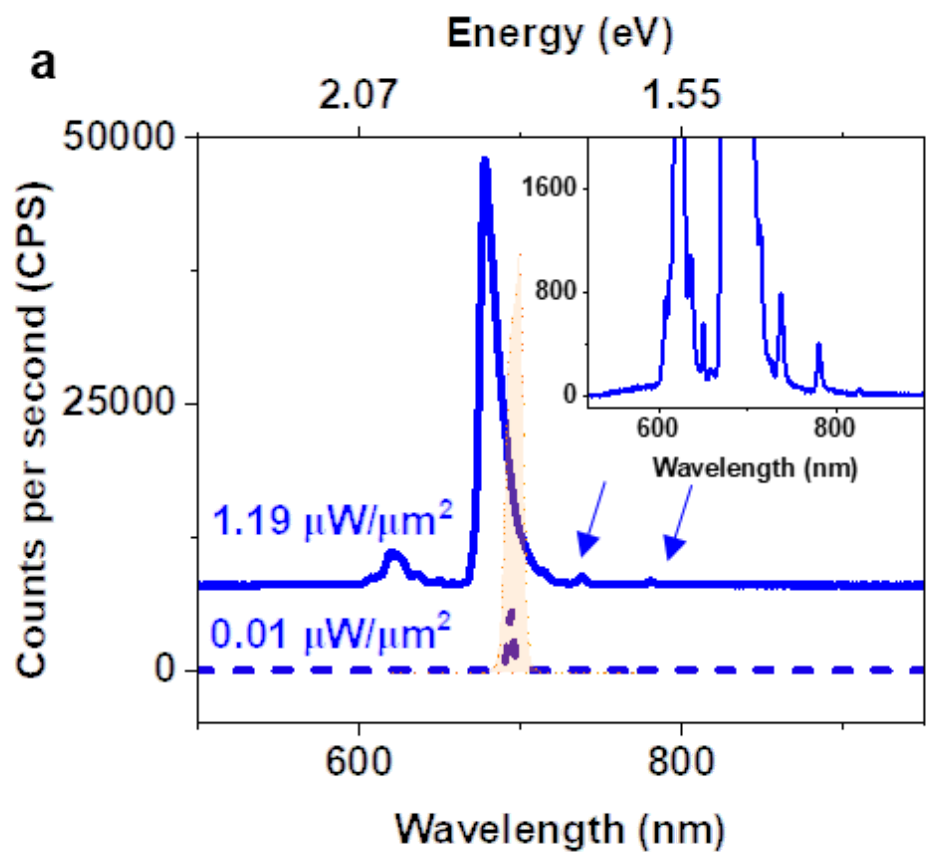

**Supplementary Figure S7.** Comparison of PL spectra (measured at T=3.5 K) collected on tent A excited with a low and high incident power densities of 10 nW/μ$m^2$ and 1.19 μW/μ$m^2$, respectively. The orange shaded area indicates the spectral window of the narrow band pass filter for photon antibunching measurements.

Section 5. Supplementary DFT simulation results of Figure 4 in the main text

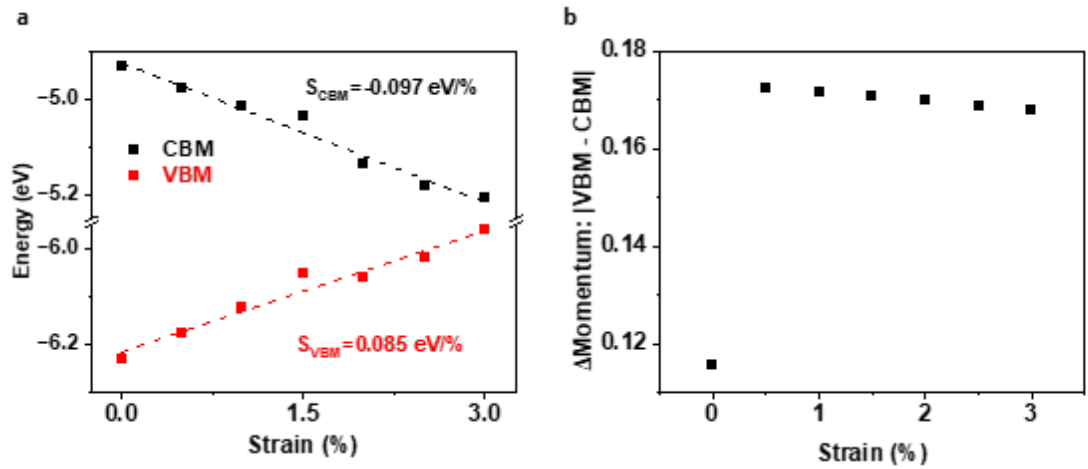

**Supplementary Figure S8. Supplementary DFT simulation results of Figure 4 in the main text. (a)** Biaxial tensile strain-dependent conduction band minimum (CBM) and valence band maximum (VBM) of bulk GaSe**. (b)** Strain-dependent momentum difference between CBM and VBM.

Section 6. Power-dependent PL spectra of the 627 nm emitter shown in Figures **5a** of main text and corresponding photon antibunching measurement results.

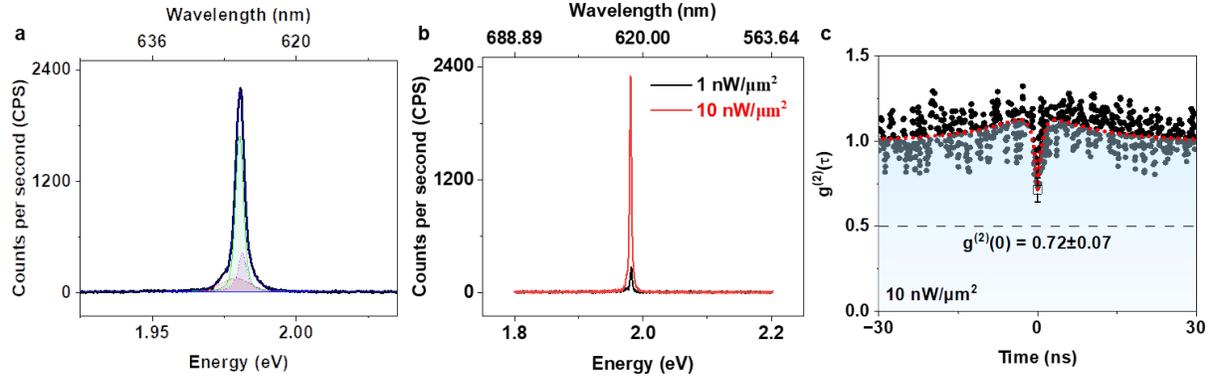

**Supplementary Figure S9. PL spectra and corresponding antibunching measurements of the 627 nm emitter (measured at T = 3.5 K, incident power density = 1 nW/$\mu m^2$ and 10 nW/$\mu m^2$) shown in the CL spectra in Figure 5a of the main text.** (a) Deconvolution of a PL spectrum of this 627 nm (1.978 eV) SPE measured at 600 g/mm grating with 0.5 meV resolution and incident power density = 10 nW/$\mu m^2$, which consists of an exciton peak at 1.980 eV (626.2 nm) and two biexciton peaks at 1.979 eV (626.6 nm) and 1.981 eV (625.9 nm). (b) PL spectra of this 627 nm emitter measured with incident power density = 1 nW/$\mu m^2$ and 10 nW/$\mu m^2$, respectively. (c) Photon-antibunching measured with incident power density = 10 nW/$\mu m^2$ and a narrow bandpass filter (627 nm, full width at half maximum (FWHM) = 10 nm); the fitted second-order correlation function shows $g^2(0) = 0.72 \pm 0.07$.

The peak at 627 nm can be deconvolved into an exciton peak at 1.980 eV (626.2 nm) and two biexciton peaks at 1.979 eV (626.6 nm) and 1.9812 eV (625.9 nm), respectively that shown in **Figure S9a**. The peak at 694 nm can be deconvolved into an exciton peak at 1.792 eV (692 nm) and a biexciton peak at 1.787 eV (694 nm) as described and identified by power-dependent measurements shown in supplementary **Figures S9a** and **S9b**. Note that these biexciton features compromise the single photon purity of related SPEs because they are energetically close to the exciton PL and hard to spectrally filter with suitable efficiency and contrast.[7]

Section 7. Power-dependent PL spectra of the 694 nm SPE shown in Figures **5e** of main text

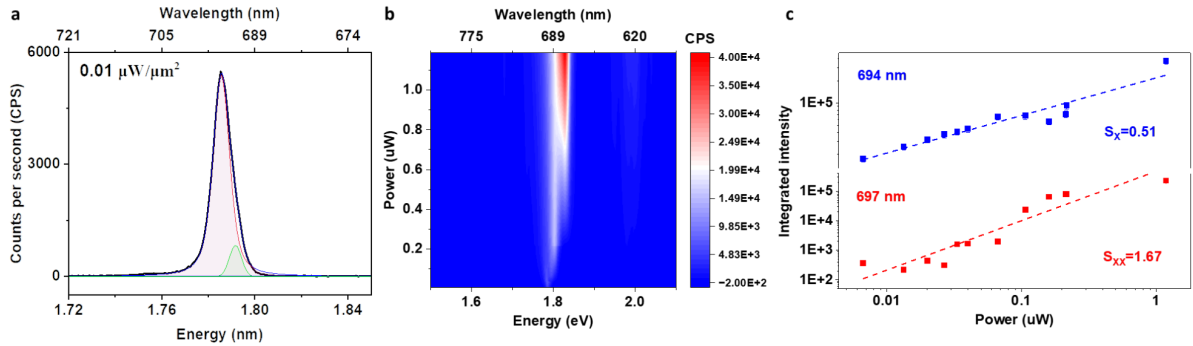

**Supplementary Figure S10. Power-dependent PL spectra of the 694 nm (1.787 eV) SPE shown in Figures 5e of main text.** (a) Deconvolution of a PL spectrum of this 694 nm (1.787 eV) SPE measured at 600 g/mm grating with 0.5 meV resolution, which consists of an exciton peak at 1.786 eV (694 nm) and a biexciton peak at 1.792 eV (692 nm), respectively. (b) Power-dependent PL spectra of this 694 nm (1.787 eV) SPE. (c) The integrated counts of the two biexcitons and the exciton as a function of incident laser power density.

Section 8. Power-dependent photon-antibunching results of the 694 nm SPE shown in Figures **5e** of main text

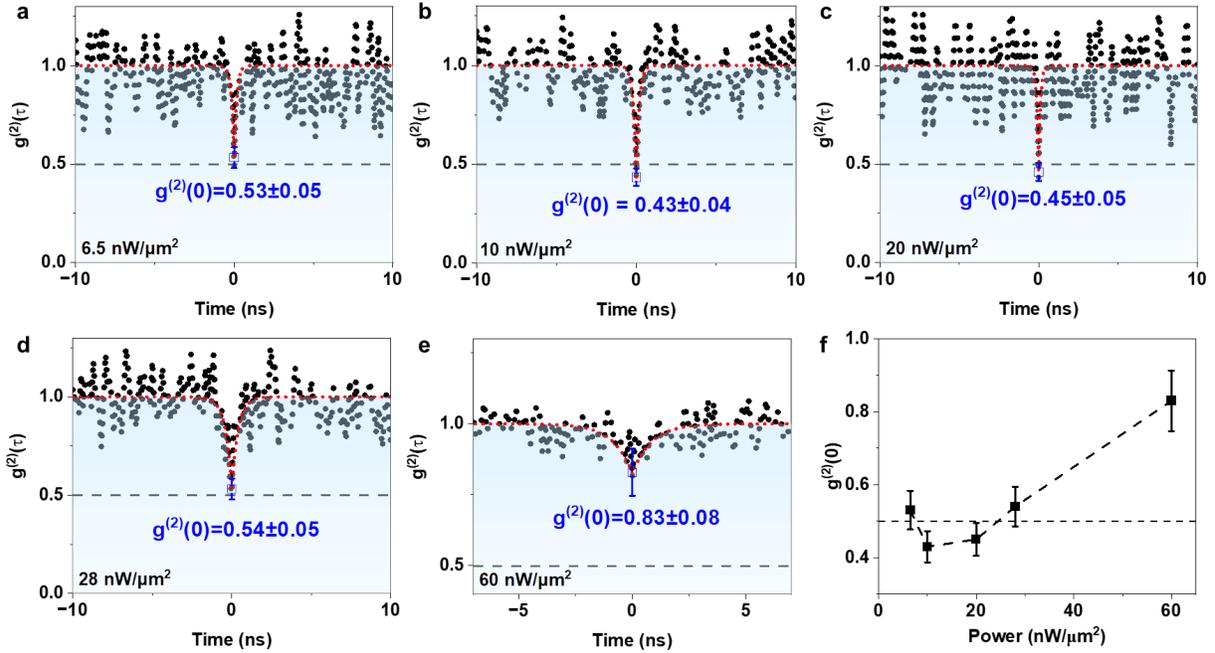

**Supplementary Figure S11. Power-dependent photon-antibunching results of the 694 nm (1.787 eV) SPE shown in Figures 5e of main text. a-e** correspond to the incident power densities of 6.5 nW/$\mu m^2$, 10 nW/$\mu m^2$, 20 nW/$\mu m^2$, 28 nW/$\mu m^2$ and 60 nW/$\mu m^2$, respectively. **f** Summary of power-dependent $g^2(0)$ values.

The photon-antibunching results are described by the second-order correlation function (intensity auto-correlation)[8]:

$$g^{(2)}(\tau) = \frac{\langle I(\tau)I(t+\tau)\rangle}{\langle I(\tau)\rangle^2}, \quad (1)$$

where $I$ represent the emission intensity, and $\tau$ is the time delay. In practice, the intensity autocorrelation function could be written as:

$$g^{(2)}(\tau) = 1 - a * \exp^{-\frac{|\tau|}{b}}, \quad (2)$$

where (1-a) is the photon-antibunching value at the zero-time delay ($g^{(2)}(0)$) and b is the antibunching decay time.

The photon bunching results are described by the second-order correlation function (intensity auto-correlation):[9,10]

$$g^{(2)}(\tau) = 1 + \rho^2 \cdot \exp^{-\frac{2\cdot|\tau|}{\tau_c}}$$, where $\rho^2$ is the degree of bunching and $\tau_c$ is the characteristic bunching (coherence) time constant.